\documentstyle[11pt]{article}
\def\sign{{\rm sign}\;}
\def\const{{\rm const}}

\def\e{{\rm e}}
\def\to{\rightarrow}
\def\da{\delta\alpha}
\def\dg{\delta\gamma}
\def\df{\delta\varphi}
\def\dn{\delta\nu}
\def\dps{\delta\psi}
\def\dW{\delta W}
\def\g{\hat g}
\def\gg{\overline{g}}
\def\ds{d \hat s}
\def\C{\hat C}
\def\L{{\cal L}}
\def\CC{\overline{C}}
\def\half{\frac{1}{2}}
\def\sumi{\sum_{i=1}^{n}}
\def\summ{\sum_{i=2}^{n}}
\def\lst{\lambda_{\rm string}}
\def\umx{u_{\max}}
\def\nn{\nonumber}
\newcommand{\vars}[1]{\left\{\begin{array}{ll}#1\end{array}\right.}

\begin{document}
\thispagestyle{empty}

\begin{center}
               RUSSIAN GRAVITATIONAL ASSOCIATION\\
               CENTER FOR SURFACE AND VACUUM RESEARCH\\
               DEPARTMENT OF FUNDAMENTAL INTERACTIONS AND METROLOGY\\
\end{center}
\vskip 4ex
\begin{flushright}                 RGA-CSVR-010/94\\
                                   gr-qc/9407033
\end{flushright}
\vskip 15mm

\begin{center}
{\large\bf EXTRA DIMENSIONS AND POSSIBLE\\
\vskip1ex
           SPACE-TIME SIGNATURE CHANGES}\\
\vskip2.5ex
     {\bf K.A.Bronnikov}\\
\vskip 5mm
     {\em Center for Surface and Vacuum Research,\\
     8 Kravchenko str., Moscow 117331, Russia}\\
     e-mail: bron@cvsi.uucp.free.net\\
\vskip 10mm
\end{center}

{\bf ABSTRACT}

\bigskip
\noindent
     Exact static, spherically symmetric solutions to the
     Einstein-Maxwell-scalar equations, with a dilatonic-type scalar-vector
     coupling, in $D$-dimensional gravity with a chain of $n$ Ricci-flat
     internal spaces are considered, with the Maxwell field potential having
        two nonzero components: the temporal, Coulomb-like one and the one
        pointing to one of the extra dimensions.
        The properties and special cases of the solutions are discussed, in
        particular, those when there are horizons in the space-time. Two types
        of horizons are distinguished: the conventional black-hole (BH) ones
and
        those at which the physical section of the space-time changes its
        signature (the latter are called {\it T-horizons}). Two theorems are
        proved, one fixing the BH- and T-horizon existence conditions, the
other
        claiming that the system under study cannot have a regular center. The
        stability of a selected family of solutions under spherically symmetric
        perturbations is studied. It is shown that only black-hole solutions
are
        stable, while all others, in particular, those with T-horizons are
        unstable.

\vskip 45mm

\centerline{Moscow 1994}
\pagebreak

\setcounter{page}{1}

\section {Introduction}

     If we recognize the highly probable reality of extra dimensions as
     suggested by modern unification theories (e.g., \cite{green}), tracing
     back to the famous works of Kaluza and Klein \cite{kaluza,klein}, it is
     natural to seek their visible manifestations in our 4-dimensional
     physics. Such predictions do exist in many works and include possible
     fundamental constant instabilities \cite{mel}, modified properties of the
     gravitational and electromagnetic fields \cite{bm-itogi} and the
     appearance of new peculiar objects (maybe possible dark matter
     candidates \cite{wesson}).

     Here we discuss a potentially observable effect able to appear only in
     multidimensional theory, namely, possible space-time signature changes.
     This happens when a horizon appears not in the two-dimensional subspace
     $(t,u)$ where $t$ is the time coordinate and $u$ a spatial one,
     belonging to the conventional physical space, but in a subspace $(u,v)$
     where $v$ belongs to one of the extra dimensions.  The coordinates $u$
        and $t$ are both time-like beyond such a horizon. Such space-time
        domains, called {\sl T-holes,} do exist in some spherically
        symmetric solutions of the multidimensional Einstein equations which
        describe gravity with reasonable accuracy in the low-energy limit of
the
        known string and supergravity theories.  Moreover, as we shall see,
        T-holes appear as solutions to multidimensional field equations as
        frequently as do black holes (BH's), therefore their possible
        observability and relevance for the physics of the early Universe and
        superdense astrophysical objects is worth studying.

     To the author's knowledge, the only space-time signature
     changes considered to-date at the classical level (see
     \cite{dray,zhuk} and references therein) are transitions between
     Euclidean and Loretzian domains, mainly in connection with models of
     the ``birth of the Universe from nothing'' \cite{vilenkin].

     We will consider exact static, spherically symmetric solutions
     to the Einstein-Maxwell-scalar equations, with a dilatonic type coupling
     between the scalar and vector fields, in $D$-dimensional gravity with a
     chain of $n$ Ricci-flat internal spaces and an arbitrary dimension
     of coordinate spheres (the group of spatial motions is $O(d+2)$), in
        the spirit of Tangherlini's generalization of the Schwarzschild metric
     \cite{tan}.

     We start from the action
\begin{equation}                                                       
     S= \int d^D x \sqrt{^D g}\Big[^D R+g^{MN}\varphi_{,M}\varphi_{,N}
               -\e^{2\lambda \varphi}F^2\Big]              \label{Action}
\end{equation}
     for the set of interacting fields of the type that arises in the field
     limit of superstring theories. Here $g_{MN}$ is the $D$-dimensional
     metric,  $^D g=\mid \det g_{MN}\mid$,\  $\varphi$ is a dilaton scalar
     field and $F^2 = F^a F^a = F^{a\,MN}F^a_{MN},\
     F^a = dW^a,\ W^a\ (a= 1,\ 2, \ldots)$ being Abelian gauge fields of
     which one is to be interpreted as the electromagnetic field. As pointed
     out in \cite{myers}, an electromagnetic field introduced in the
     multidimensional action may seem less aesthetic than a purely
     gravitational (Einstein) action frequently considered in Kaluza-Klein
     theories but it appears that elementary gauge fields are necessary to
     obtain a realistic grand unification theory. Two types of $W^a$ will be
     considered: $W^1$, the Coulomb-like one, for which the vector potential
     is $t$-directed, and $W^2$ pointing to one of the extra dimensions.

     The field-theoretic limit of string
     theory corresponds to the specific value of the coupling constant
     $\lambda = \lst = \pm (D-2)^{-1/2}$ \cite{green,shira}.

     The field equations are written down in Sect. 2 and solved in Sect. 3.
     We come through a striking coincidence: when both $W^1$ and $W^2$ are
     nontrivial, the field equations are explicitly integrable if and only
     if $\lambda^2 = 1/(D-2)$, i.e., exactly for the dilaton coupling which
     follows from string theory.

     In Sect. 4 special cases are discussed, with references to the results
     obtained earlier and appropriate coordinate transformations. In Section
     5 we indicate the special cases when the solutions exhibit BH or
     T-hole horizons and prove a theorem telling us that if both vector
     components $W^1$ and $W^2$ are nonzero, then no horizons are possible,
     whatever the coupling $\lambda$ is. Another theorem is also proved, that
        the system under study cannot have a regular center.

     Sects. 6 and 7 discuss different conformal gauges in
     $D$ and 4 dimensions, respectively. The point is that if the underlying
        theory is string theory, then a more fundamental role is played by the
        ``string metric'', or ``$\sigma$ model metric''
        $\g_{AB}=\e^{-2\lambda\varphi}g_{AB}$ rather than $g_{AB}$ from
        \ref{Action} (see, e.g., \cite{shira,banks} and references therein).
        Although mathematically a transition from $\g_{AB}$ to $g_{AB}$ may be
        treated as a change of variables simplifying the field equations,
        such issues as the nature of singularities (if any) and topology are
        better to discuss in terms of $\g_{AB}$. (Strictly speaking, this
        argument does not apply to $\lambda \ne \lambda_{\rm string}$ when the
        underlying more fundamental theory is not definitely fixed).

     On the other hand, the observable effects in 4 dimensions depend on how
     nongravitational matter interacts with the metric and dilaton fields and
     are described in different ways in different ``conformal gauges'', or
     systems of measurement, which are discussed in Sect. 7. It should be
        stressed that such things as horizons and signatures are the same in
all
        the relevant conformal gauges since the conformal factors connecting
        them are regular at the horizons.

     Section 8 is dedicated to a stability
     investigation under monopole perturbations preserving spherical
     symmetry. It is found, in agreement with the previous results,
     that only BH solutions are stable under such perturbations,
     while all others, in particular, those with T-horizons, are unstable.
     Section 9 contains some concluding remarks.

     Throughout the paper capital Latin indices range from 0 to $D-1$, Greek
     ones from 0 to $d+2$ and the index $i$ enumerates subspaces.

\section {Field equations}

     The set of field equations corresponding to (\ref{Action}) is
\begin{eqnarray}
 \nabla^M \nabla_M \varphi +\lambda \e^{2\lambda\varphi}F^2&=&0,
                                                            \label{Ephi}\\
         \ \ \nabla_N(\e^{2\lambda\varphi}F^{a\,NM})&=&0,\label{EMax}\\
     R_{MN}-g_{MN}R^A_A/2 &=& -T_{MN}                      \label{Einst}
\end{eqnarray}
     where $T_{MN}$ is the energy-momentum tensor
\begin{equation}
     T_{MN} = \varphi_M\varphi_N - \half g_{MN}\varphi^A\varphi_A +
    \e^{2\lambda\varphi}
     \big[-2F_M^{a\ A}F^a_{NA}+\half g_{MN}F^2\big].            \label{EMT}
\end{equation}

     Consider a $D$-dimensional Riemannian or pseudo-Riemannian manifold
     $V^D$ of the structure
\begin{equation}
     V^D = M^{(d+3)} \times M_1\times \ldots
     \times M_n; \ \ \dim M_i=N_i; \ \ D=3+d+\sumi N_i,
                                                              \label{Stru}
\end{equation}
     where $M^{(3+d)} = M^2 \times S^{d+1}$
     plays the role of the conventional space-time and $M_i$
     are Ricci-flat manifolds of arbitrary dimensions and signatures
     with the line elements
     $ds_i^2$, $i=1,\cdots, n$. We seek solutions of the field equations
     such that $M^{(3+d)}$ is a static,
     (hyper)spherically symmetric space-time with the metric
\begin{equation}
     ds_{d+3}^2 =
     \e^{2\gamma(u)}dt^2-\e^{2\alpha(u)}du^2 - \e^{2\beta(u)}d\Omega_{d+1}^2
                                                                \label{Ds4}
\end{equation}
     where $d\Omega_{d+1}^2$ is the line element on a unit
     $(d+1)$-dimensional sphere $ S^{d+1}$,
     while all the scale factors $\e^{\beta_i}$ of the internal spaces
     $M_i$ depend on the radial coordinate $u$, i.e., the $D$-metric is
\begin{equation}
       ds_D^2 = g_{MN}dx^M dx^N=
          ds_{d+3}^2 + \sumi \e^{2\beta_i(u)}ds_i^2.        \label{DsD}
\end{equation}
     If we denote $\gamma =\beta_{-1}, N_{-1}=1,  \beta  =\beta_0,  N_0=d+1$
     and choose the harmonic radial coordinate $u$ \cite{br-acta} such that
\begin{equation}
     \alpha = \sum_{i=-1}^{n}N_i \beta_i,                       \label{Harm}
\end{equation}
     the Ricci tensor components $R_M^N$ can be written in the highly
     symmetric form ($x^1=u$)
\begin{eqnarray}
     R_u^u &=& -\e^{-2\alpha}\sum_{i=-1}^{n}N_i [\beta_i''
       +\beta_i^{'2}-\beta_i'\alpha']; \nn \\
     R_{\mu}^N &=& 0 \ \ \ (N>d+2;\ \ \mu =0,\cdots ,d+2); \nn \\
     R_{a_j}^{b_i} &=&
     -\delta_j^i\delta_{a_i}^{b_i}\e^{-2\alpha}\beta_i'', \ \
                    i\ne 0                                  \nn\\
     R_2^2 &=& \ldots= R_{d+2}^{d+2} = (d+1)\e^{-2\beta}
          -\e^{-2\alpha}\beta'';                               \label{Ricci}
\end{eqnarray}
     where primes denote $d/du$ and the indices
     $a_j\ (b_i)$ refer to the subspace $M_j\ (M_i)$.

     The fields $\varphi$ and $F$ compatible with the assumed symmetry
     include $\varphi=\varphi (u)$, a component of $W^a$ in the $t$
     direction (the Coulomb electric field, $W^1 = U(u)dt$), similar
     components in some internal one-dimensional subspaces (if any), so that
     $W^2 = W(u)dv$, etc. ($v$ is the corresponding internal coordinate)
     and, provided $d=1$, a monopole magnetic field of the form
     $W_m = q_m \cos\theta\, d\phi$ where $q_m$ is the magnetic charge,
     $\theta$ and $\phi$ are the conventional spherical coordinates.

     In $D$ dimensions a magnetic charge is not included into the
        solution in such a simple way as
     it is for $D=4$, especially when $\lambda=0$. This wiil be the
     subject of a subsequent paper; here we would like to consider a system
     with two gauge field components $W^1$ (Coulomb) and $W^2$ (as
        described above; the corresponding subspace will be denoted $V_1$, with
        $N_1=1,\ \beta_1=\nu (u)$).  So the gauge field strengths are
$$
     F^a = F^a_{AB}dx^A\wedge dx^B;\ \
     F^1 = U'(u)\,du\wedge dt;\ \ F^2=W'(u)\,du\wedge dv.
$$
     The vector field equations lead to
\begin{equation}
     \e^{2\lambda\varphi+2\alpha}F^{1\,ut} = q = \const,\ \
     \e^{2\lambda\varphi+2\alpha}F^{2\,uv} = q'= \const.         \label{Maxw}
\end{equation}
     Here $q$ is the ordinary electric charge and $q'$ is the
     gauge charge connected with $F^2_{uv}$.

     Now the scalar field equation and some linear combinations of the metric
     field equations may be written in the form
\begin{eqnarray}
     \varphi''+2\lambda\Big[q^2\e^{2\omega}
               + \eta_v{q'}^2\e^{2\psi}\Big]&=& 0;          \label{Ephi1}\\
     (N+1)\gamma''-2N q^2\e^{2\omega} + 2\eta_v{q'}^2\e^{2\psi}&=& 0;
                                                            \label{Egamma}\\
     (N+1)\nu''+ 2q^2\e^{2\omega} -2N\eta_v{q'}^2\e^{2\psi}&=& 0;
                                                            \label{Enu}\\
     (N-1)\beta_i''+\gamma''+\nu''&=&0,
                \ \ i=2,\ldots, n;                          \label{Ebetai} \\
     (\alpha-\beta)''-d^2 \e^{2\alpha-2\beta} &=&0;          \label{Ebeta}\\
     {\alpha'}^2-\sum_{i=-1}^{n}N_i{\beta_i'}^2 -d(d+1)\e^{2\alpha-2\beta}&=&
     \varphi'^2 - 2q^2\e^{2\omega}-2\eta_v {q'}^2\e^{2\psi}      \label{Int}
\end{eqnarray}
     where we have denoted
\begin{equation}
     \omega= \gamma-\lambda\varphi;\ \ \ \psi= \nu-\lambda\varphi;\ \
                        \ \  N = D-3.                         \label{Defopsi}
\end{equation}
     Eq.(\ref{Int}) is the $(^u_u)$ component of the Einstein equations and
     represents a first integral of (\ref{Ephi1}-\ref{Ebeta}).

\section{Exact static solutions}

     Eqs.(\ref{Ebetai},\ref{Ebeta})  are easily integrated to give
\begin{eqnarray}
 \beta_i &=&-(\gamma+\nu)/(N-1) + h_i u,\ \
          h_i ={\rm const,}\ \ i=2,\ldots,n;             \label{Betai}\\
    \e^{\beta-\alpha} &=& d\cdot s(k,u), \ \ \ \ \ \ \ \ \ k ={\rm const}
                                                            \label{Beta}
\end{eqnarray}
     where for any variables by definition
\begin{equation}
     s(c,x) = \vars{
     c^{-1}\sinh\,cx, \ & c>0; \\
     x, &  c=0;  \\
     c^{-1}\sin\,cx,  & c<0   }                             \label{Def-s}
\end{equation}
     and inessential integration constants
     have been eliminated by shifting the origin of $u$ and rescaling the
     coordinates in the subspaces $M_i$ (their true scales are thus hidden
     in $ds_i^2$).

     By our choice of the origin of $u$, with no loss of generality it can be
     asserted that the harmonic $u$ coordinate is defined for $u>0$ and $u=0$
     corresponds to spatial infinity. By (\ref{Beta}), at the asymptotic
     $u\to\infty$ the conventional flat-space (hyper)spherical radial
     coordinate $r=\e^{\beta}$ is connected with $u$ by $d\cdot u = 1/r^d$.

     From the remaining equations (\ref{Ephi1}--\ref{Enu}) the field
     $\varphi$ may be expressed in terms of the combinations (\ref{Defopsi})
     which obey the equations
\begin{eqnarray}
     (N+1)\omega'' &=& 2(N+a^2)q^2\e^{2\omega}+                         
                         2(a^2 -1)\eta_v{q'}^2\e^{2\psi};    \label{Eomega}\\
     (N+1)\psi'' &=& 2(a^2-1 )q^2\e^{2\omega}+                         
                         2(N+a^2)\eta_v{q'}^2\e^{2\psi}      \label{Epsi}
\end{eqnarray}
     where $a^2 = \lambda^2 (N+1)$. This set of equations may be integrated
     in the following three cases:
\begin{description}
\item[A.]   the conventional case: $W^1 = U(u)dt;\ q'=0,\ W^2=0$.
\item[B.]   $q=0,\ W^1=0;\ W^2 = W(u)dv$: no Coulomb-like electric field
        but the extra-dimensional vector field is nonzero;
\item[C.]   $a^2=1$, or $\lambda^2=1/(N+1)$, the special choice of the
     coupling constant, exactly the one corresponding to the strings.
\end{description}

\vskip0.5ex
     {\bf Solution A} \cite{potsdam,br-vuz91,bm-itogi}
\vskip0.5ex
\nopagebreak

     Eqs. (\ref{Ephi1},\ref{Egamma}) and (\ref{Eomega}) yield
\begin{eqnarray}
 \beta_i &=&-\gamma/N + h_i u,\ \
          h_i ={\rm const,}\ \ i=1,\ldots,n;             \label{BetaiA}\\
     \varphi &=& Cu/A -2\lambda N_{+}\omega,              \label{PhiA}\\
     \gamma &=& (\omega +\lambda Cu)/A,                   \label{GammaA}\\
     \e^{-\omega} &=& Qs(h,u+u_1); \ \ h,u_1={\rm const};\ \ Qs(h,u_1)=1
                                                          \label{OmegaA}
\end{eqnarray}
     where  $h$ and $C$ are integration constants; other
     constants are defined by
\begin{equation}
     A = 1+\lambda^2(N+1)/N, \ \ Q^2=q^2/N_{+},\ \
          N_+=(N+1)/(2AN).                               \label{Defconst}
\end{equation}
     The last condition from (\ref{OmegaA}) is the requirement that $\gamma
     =0$ at infinity, i.e., $dt$ is a time interval measured by a distant
     observer at rest with respect to our static configuration.  It should be
     noted that $\nu\equiv\beta_1$ is now indistinguishable among other scale
     factors $\beta_i$.

     The final form of the $D$-dimensional metric is
\begin{equation}
     ds_D^2=\e^{2\gamma}dt^2 -
 \e^{-2\gamma/N}\left\{\left[\frac{\e^{-Bu}}{d\cdot s(k,u)}\right]^{2/d}
     \left[\frac{du^2}{d^2 s^2 (k,u)}+d\Omega_{d+1}^2\right]
     -\sumi \e^{2h_i u}ds_i^2\right\},                   \label{DsA}
\end{equation}
     where $\gamma$ is determined by (\ref{GammaA}), the fields $\varphi$ and
     $F$ by (\ref{PhiA}) and (\ref{Maxw}) with $q'=0$,
     $B=\sumi N_i h_i$ and
     the integration constants are connected by the relation due to
     (\ref{Int})
\begin{equation}
     \frac{d+1}{d}k^2\sign k = 2N_{+}h^2 \sign h +
     \frac{C^2}{A}+\frac{B^2}{d}+\sumi N_i h_i^2.                \label{IntA}
\end{equation}
     This is exactly the solution obtained in \cite{potsdam} and, for $d=1$,
     in \cite{br-vuz91}.

\vskip 1ex
     {\bf Solution B}
\nopagebreak
\vskip 1ex

     The solution can be found by proper substitutions in that of Case
     A due to the symmetry between the variables $t$ and $v$.
        Namely, we obtain:
\begin{eqnarray}
     ds_D^2&=&\eta_v\e^{2\nu}dv^2                             \nn \\
     &+& \e^{2\nu/N}
     \left\{\e^{2h_0 u}dt^2
     -\left[\frac{\e^{-Bu}}{d\cdot s(k,u)}\right]^{2/d}
     \left[\frac{du^2}{d^2 s^2 (k,u)}+d\Omega_{d+1}^2\right]
     +\summ \e^{2h_i u}ds_i^2\right\},                   \label{DsB} \\
     \varphi &=& Cu/A -2\lambda N_{+}\psi,               \label{PhiB}\\
     \nu &=& (\psi +\lambda Cu)/A,                       \label{NuB}\\
     \e^{-\psi}&=&\vars{
                         s(h', u+u_2)\ \ &{\rm if}\ \ \eta_v=+1,\\
                   (1/h')\cosh h'(u+u_2) &{\rm if}\ \ \eta_v=-1,}
                                                        \label{PsiB}
\end{eqnarray}
     with ${Q'}^2 = {q'}^2/N_+$;
     the integration constants $C,\ h',\ h_0,\ h_i,\ k$ and $u_2$ are
     constrained by the conditions $\psi(0)=0$ and
\begin{equation}
     \frac{d+1}{d}k^2\sign k = 2N_{+}{h'}^2 \sign h' +
     \frac{C^2}{A}+\frac{B^2}{d}+h_0^2+\summ N_i h_i^2.      \label{IntB}
\end{equation}
     In (\ref{DsB}) and (\ref{IntB}) $B$ is expressed as
     $B=h_0+\summ N_ih_i$.

\vskip0.5ex
     {\bf Solution C}
\nopagebreak
\vskip0.5ex

     Eqs. (\ref{Eomega}) and (\ref{Epsi}) are easily integrated and after the
     corresponding substitutions we get:
\begin{eqnarray}
     \e^{-\omega}&=&\sqrt{2q^2}s(h, u+u_1);                      \label{OC}\\
     \e^{-\psi} &=&\left\{ \begin{array}{ll}
                 \sqrt{2{q'}^2} s(h', u+u_2)\ \ &{\rm if}\ \ \eta_v=+1,\\
                (\sqrt{2{q'}^2}/h')\cosh h'(u+u_2) &{\rm if}\ \ \eta_v=-1.
     \end{array}\right.                                       \label{PsiC}\\
     (N+1)\lambda\varphi &=& C'u -\omega -\psi;            \label{PhiC}\\
            (N+1) \gamma &=& C'u + N\omega -\psi;         \label{GammaC}\\
            (N+1) \nu    &=& C'u -\omega + N\psi;         \label{NuC}\\
            (N+1)\beta_i &=& [h_i-2C'/(N-1)] u -\omega-\psi,\label{BetaiC}
\end{eqnarray}
     and the $D$-dimensional metric has the form

\bigskip
$  ds_D^2 = \eta_v\e^{2\nu}dv^2 + \e^{2\gamma}dt^2 $
\begin{equation}
+\exp\bigg[-\frac{4C'u}{N^2-1}-\frac{2(\omega+\psi)}{N+1}\bigg]
            \bigg\{-\bigg[\frac{\e^{-B'u}}{d\cdot s(k,u)}\bigg]^{2/d}
            \left[\frac{du^2}{d^2 s^2 (k,u)}+d\Omega_{d+1}^2\right]
               +\summ \e^{2h_i u} ds_i^2 \bigg\}                \label{DsC}
\end{equation}
     where $B'= \summ N_i h_i$.
     The relation due to (\ref{Int}) among the constants is
\begin{equation}
     \frac{d+1}{d}k^2\sign k = h^2 \sign h + {h'}^2\sign h'
    + \frac{{C'}^2}{N-1}+\frac{{B'}^2}{d}+\sum_{i=2}^{n} N_i h_i^2.
                                                                \label{IntC}
\end{equation}
\medskip

     The static, spherically-symmetric solutions have in general $(n+4)$
     essential integration constants: the scalar charge $C$ or $C'$, the
     ``electric'' charges $q$ and $q'$, $n$ ``charges'' of extra dimensions
     $h',\ h_i$, and also $h$ and $k$; these $(n+5)$ constants are connected
     by (\ref{IntA}) or (\ref{IntC}). In Cases A and B one of the charges $q$
     or $q'$ is zero and the number of independent constants is $n+3$.

\section{Special cases}

     Let us point out some special cases of the solutions.
\begin{description}
\item[(a)]
     The gauge field $F$ is eliminated when $q=q'=0$. Then from
     either of the solutions A, B, C the generalized Tangherlini solution
     \cite{fim} is recovered (the transition is possible for $k>0,\ h>0$ and
     (or) $h'>0,\ u_1>0$).  Thus, the transformation
\begin{equation}
     \e^{-2ku}= 1-\frac{2k}{R} \equiv f(R);
      \ \ \frac{-h+\lambda C}{A}=ka_0;
      \ \ h_i=k\left(-a_i+\frac{a_0}{N}\right),               \label{e-2ku}
\end{equation}
     applied to Solution A brings it to the form
\begin{eqnarray}
     ds_D^2 &=& f^{a_0}dt^2 -[R^2f^{1-a_0-b}]^{1/d}\left[\frac{dR^2}{R^2 f}+
               d\Omega_{d+1}^2\right]
               + \sumi f^{a_i}ds_i^2, \nn \\
       \varphi &=&-\frac{\CC}{2k}\ln f(R), \ \
               \CC=\frac{1}{A}(C+2\lambda N_{+}h), \nn \\
       (d+1)/d &=& \sum N_i a_i^2+a_0^2+ (a_0+b)^2/d + \CC^2/k^2
                                                            \label{GTan}
\end{eqnarray}
     where $b = \sumi N_i a_i$. If all $h_i=0$, the 4-dimensional part of the
     solution (assuming $d=1$) coincides with
     that known for a minimally coupled scalar field in general relativity
     \cite{fish}, though in other notation. Furthermore, if $\CC=0$,
     the  Schwarzschild solution is recovered.  The original Tangherlini
     solution \cite{tan} corresponds to the special case $\CC=0,\
     n=0$, i.e., when there is no scalar field and no internal spaces, so
     that $D=d+3,\ N=d$. Peculiar features of the solution (\ref{GTan}) with
        $\CC=0$ and many temporal coordinates are discussed in \cite{ivmel}.
        Multitemporal space-times were earlier discussed, in particular, by
        A.D.Sakharov \cite{sahar}

\end{description}

     The scalar field in (\ref{GTan}) affects the metric only through the
     constant $\CC$ in the relation among the constants.

     The space-time (\ref{GTan}) has a horizon at
     $r=2k$ only in the simplest case $ a-1 = a_i =b = \CC = 0$.
     Thus in this model an electrically neutral BH exists only with
     ``frozen'' extra dimensions ($\e^{\beta_i}={\rm const}$) and with no
     scalar field.  (The latter result is well known in conventional general
     relativity as one of the no-hair theorems: a massless, minimally coupled
     scalar field is incompatible with an event horizon). In this sense the
     generalized Schwarzschild-Tangherlini BH's may be called trivial.

\begin{description}
\item[(b)]
     When $\lambda =0$, Solution A reduces to the generalized
     Reissner-Nordstrom-Tangherlini solution for linear scalar and
     electromagnetic fields (\cite{br-ann} for $d=1$);
     at $d=1,\ n=0$ it coincides with the Penney
     solution \cite{penney} which in turn reduces to the Reissner-Nordstrom
     (RN) one when the scalar field is eliminated.
\item[(c)]
     The scalar field is ``switched off'' when $\lambda =C=0$.
     Then for Solution A we arrive at a special case of item (b) discussed in
     Ref.\cite{fim}. It again reduces to the RN solution when $d=1,\ n=0$.
\end{description}
     A new feature implied by a nonzero electric charge as compared with item
     (a) is that the constants $k,\ h$ and (or) $h'$ can have either sign and
     the function $s(h, u+u_1)$ can be sinusoidal, which yields
     $u_{\max}<\infty$. Physically that means the appearance of a
     RN-like repelling singularity at the center of the
     configuration.
\begin{description}
\item[(d)]
     When in Case A extra dimensions are absent, we arrive at a solution for
     interacting scalar and electromagnetic fields in $(d+3)$-dimensional
     general relativity; for $d=1$ it was first obtained in
     Ref.\cite{brsh-vuz}.
\item[(e)]
     For the case of 4-dimensional physical space-time ($d=1$, i.e.,
     the space $M^{3+d}=M^4 = R^1 \times R^1 \times S^2$),
     Solution A was obtained in \cite{br-vuz91}.
\item[(f)]
     The case $D=5\ (d=n=N_1=1)$ with no dilaton and gauge fields was
     considered in Refs.  \cite{kramer,davow} and many subsequent papers; it
     also coincides with the Kaluza-Klein soliton considered by Gross and
     Perry \cite{groper}. Similar solutions for $D=6$ and $D=7$ are presented
     in \cite{vladim}; see also references therein.
\item[(g)]
     There are special cases of Solutions A and B when the space-time
     exhibits horizons. They are discussed in the next section.
\end{description}

\section{Horizons: black holes and time holes}

\noindent{\bf 5.1.}
     The behavior of the metric coefficients
     for different combinations of integration constants is rather various.
     However, calculations show that Solution A has a naked
     singularity at $u=u_{\max}$ or $u=\infty$ in all cases except
\begin{equation}
     h_i=-k/N; \ \ \ h=k>0; \ \ \ C=-\lambda k(N+1)/N,       \label{ConBH}
\end{equation}
     when the sphere $u=\infty$ is a Schwarzschild-like event horizon:
     at finite radius $r=\e^{\beta}$ of a coordinate sphere
     the metric coefficient $g_{tt}=0$ and the light travel
     time $\int \exp(\alpha -\gamma )du$ diverges.

     Although all the subsequent results may be formulated for arbitrary $d$
     (for groups of spatial motions $O(d+2)$), in what follows we restrict
     ourselves to the physical case $d=1$ in order not to overload the
     formulas. Thus hereafter $d\Omega^2=d\Omega_2^2= d\theta^2 +
        \sin^2\theta d\phi^2$.

     In the BH case (\ref{ConBH}) only two independent integration
     constants remain, say,  $k$ and $Q$, and the coordinate transformation
     from (\ref{e-2ku}) brings the solution to the form \cite{br-vuz91}
\begin{eqnarray}
     ds_D^2 &=& \frac{(1-2k/R)dt^2}{(1+p/R)^{2/A}}
                -(1+p/R)^{2/AN}\left[\frac{dR^2}{1-2k/R}
               +R^2 d\Omega^2 - \sumi ds_i^2\right];             \nn\\
     \e^{\lambda\varphi} &=& (1+p/R)^{a^2/(a^2+N)};\ \ \
     F = F^1_{Rt} dR\wedge dt, \ \ F^2_{Rt} = q/(R+p)^2.         \nn\\
     p &=& (k^2+Q^2)^{1/2} - k.                                  \label{BH}
\end{eqnarray}
     This solution \cite{br-vuz91} extends the well-known dilatonic
     BH solution (see, e.g., \cite{gibmaeda,garf,hein}) to spaces of
     the form (\ref{Stru}).  For the first time it was obtained in 4
        dimensions \cite{brsh-vuz} where for $\lambda=0$ they reduce to the RN
        solution.  The solution is brought to more frequently used notations by
        the substitutions
\begin{equation}
        R+p = r;\ \  p=r_-;\ \ 2k= r_+ - r_-.
\label{r+}
\end{equation}

     A certain combination of $k$ and $Q$ must be associated with the
     gravitating (Schwarzschild) mass of the configuration. However, its
     particular form depends on the choice of the physical metric, to be
     discussed in Sect. 7.

     In this family of BH solutions a nonzero scalar dilaton
     field exists solely due to the interaction ($\lambda\ne
     0$).  When $\lambda=0$, i.e., the $\varphi$ field becomes minimally
     coupled, a horizon is compatible only with $\varphi=$const. This
     conforms to the well-known ``no-hair'' theorems and the properties of
     the general-relativistic scalar-vacuum and scalar-electrovacuum
     configurations.

\vskip1ex
\noindent{\bf 5.2.}
     The above BH solutions have counterparts among the Class B
     solutions. Namely, under the same conditions (\ref{ConBH}) (where just
     $h$ is replaced by $h'$) the sphere $u=\infty$ is a horizon as well but
     now in the $(v,u)$ subspace instead of $(t,u)$ in the previous,
     conventional case. The solution is
\begin{eqnarray}
     ds_D^2 & = & \frac{(1-2k/R)\eta_v dv^2}{(1+p/R)^{2/A}}
          + (1+p/R)^{2/AN}\bigg[dt^2- \frac{dR^2}{1-2k/R}
          - R^2 d\Omega_2^2 + \summ ds_i^2\bigg],     \nn\\
     \e^{\lambda\varphi} &=& (1+p/R)^{a^2/(a^2+N)};\ \ \
     F = F^2_{Rv} dR\wedge dv, \ \ F^2_{Rv} = q'/(R+p)^2.        \nn\\
     p &=& (k^2+\eta_v Q'^2)^{1/2} - k.                     \label{TH}
\end{eqnarray}

     The main feature of these configurations is that the physical space-time
     $M^4$ changes its signature at $R=2k$: it is $(+---)$ at $R>2k$ and
     $(++--)$ at $R<2k$. This evidently means that the anomalous domains
     should contain quite new, unconventional physics whose possible
     consequences and observational manifestations are yet to be studied. It
     is suggested to call these domains with an unusual space-time signature
     {\sl time holes} or {\sl T-holes} and the corresponding horizons {\sl
     T-horizons}.

     Evidently at $R=2k$ the signs of $g_{uu}$ and $g_{vv}$ simultaneously
     change. Moreover, if $\eta_v =1$, i.e., this compactified direction is
     timelike at big $R$, the overall signature of $V^D$ is preserved but in
     the opposite case, $\eta_v =-1$, it is changed by four: two spacelike
     directions become timelike. However, it can be directly verified that a
     T-horizon is not a curvature singularity, either for the $D$-dimensional
     metric or for its 4-dimensional section.

     It might seem that time holes can appear only when there is a gauge
     field pointing to an extra dimension. However, putting $q'=0$ in
     (\ref{TH}), we come to a direct analog of the Schwarzschild
     solution (to be called {\it T-Schwarzschild}) which for $D=5$
     coincides with the zero dipole moment soliton in the terminology of
     \cite{groper}:
\begin{equation}
     ds_D^2 = (1-2k/R)\eta_v dv^2 + dt^2 -
     (1-2k/R)^{-1}dR^2 -R^2 d\Omega_2^2 +\sum_{i=2}^{n}ds_i^2  \label{TSchw}
\end{equation}
     while both fields $\varphi$ and $F$ are zero.

     It can be concluded that each BH configuration of any dimension
     has a family of time-hole counterparts (a family since the subspaces
     $V_i$ may have different dimensions and signatures) and {\sl vice
     versa}. However, if a BH possesses an external field, such as
     the Coulomb field of a RN BH, its T-hole analog has a field
     modified  by the $t\leftrightarrow v$ interchange, as is the case with
     the above solutions: the Coulomb-like field becomes the one pointing
        in the $v$ direction which from the 4-dimensional viewpoint looks like
a
     specific scalar field interacting with the dilaton (see Sect. 7).

     Unlike a BH horizon, a T-horizon is not in absolute past or
     future from a distant observer's viewpoint, it is visible since it takes
     a finite time for a light signal to come from it
     ($\int\e^{\alpha-\gamma}du < \infty$, independently of a conformal
     gauge).

\vskip1ex
\noindent{\bf 5.3.}
    It can be directly verified that among the Class C solutions
    (\ref{OC}-\ref{IntC}) there are no special cases with horizons, either
    BH or T-hole ones. Moreover, one can prove that there are no
    solutions with horizons in the more general case when exact analytic
    solutions are hard to obtain, namely, when both charges $q$ and $q'$ are
    nonzero and the dilaton coupling constant $\lambda$ is arbitrary.

    To do so let us adopt a convenient horizon definition for our static,
    spherically symmetric configurations. Namely, we will call a {\sl
    BH-horizon} (i.e., a conventional BH horizon) a sphere in
    a space with the metric (\ref{DsD}) where the metric
    functions $\beta$ and $\beta_i$ are finite while
    $g_{tt}= \e^{2\gamma} \to 0$. Similarly, a {\sl T-horizon} is a sphere
    where $\beta,\ \gamma,\ \beta_i\ (i=2,\ldots,n)$ are finite
    while $g_{vv}=\e^{2\nu}\to 0$ where $v$ parametrizes the one dimensional
    subspace $V_1$, one of the internal subspaces. (The BH-horizon
    definition would conform to the conventional notion of an event horizon
    in the $(t,u)$ subspace if we required that, in addition, the
    travel time $\int \e^{\alpha-\gamma}du$ of a light signal approaching
    it be infinite; however, this condition is nowhere used in our
    argument.)

\vskip 0.7ex

{\bf Theorem 1.} {\sl The static, spherically symmetric field system}
     (\ref{Action},\ref{Stru},\ref{DsD},\ref{Maxw}) {\sl has neither a
        BH-horizon, nor a T-horizon if both $q$ and $q'$ are nonzero.}

\vskip 0.5ex

{\sl Proof.} At our disposal are Eqs.(\ref{Betai}-\ref{Epsi}) and the first
     integral of Eqs. (\ref{Eomega},\ref{Epsi}) for $\omega$ and $\psi$ which
     follows from (\ref{Int}) and has the form
\begin{equation}
     -\frac{d+1}{d}k^2\sign k + \frac{{B'}^2}{d} + \summ N_i h_i^2
     +\frac{C^2}{A'} + \half \bigg[(\psi'-\omega')^2
               +\frac{N+1}{N-1}\frac{(\psi'+\omega')^2}{A'}\bigg]
     = 2q^2\e^{2\omega} + 2\eta_v  {q'}^2\e^{2\psi}             \label{IntG}
\end{equation}
     and the other variables are expressed in terms of $\omega$ and $\psi$:
\begin{eqnarray}
     \varphi &=& (Cu -\omega -\psi)/A',\ \ C=\const;      \label{PhiG}\\
     \gamma &=& \frac{1}{A'}\Big[\lambda Cu
            +\omega +\frac{a^2}{N-1}(\omega -\psi)\Big];  \label{GammaG}\\
     \nu &=& \frac{1}{A'}\Big[\lambda Cu
                +\psi - \frac{a^2}{N-1}(\omega -\psi)\Big]; \label{NuG} \\
     \beta_i &=& h_i u - (\gamma+\nu)/(N-1),\ \ h_i=\const,\
                                             i= 2,\ldots,n; \label{BetaiG}\\
     \beta &=& -\frac 1d \big[ \ln s(k,u) + B'u] - \frac{\gamma+\nu}{N-1}.
                                                            \label{BetaG}
\end{eqnarray}
     where
$$
     A' = 1+2a^2/(N-1) = 1 + 2\lambda^2(N+1)/(N-1),\ \ \
          B' = \summ N_i h_i.
$$

     Let us assume that there a BH-horizon at some $u=u^*$.

     Evidently $u^*$ cannot be finite. Indeed, if it were,
     then (\ref{BetaiG}) would immediately lead to $\beta_i (u^*)=\infty$
     since $\gamma(u^*)+\nu(u^*)=-\infty$. It is still possible that there is
     no extra dimension except $V_1$ parametrized by $v$; however, then the
     finite $\beta$ condition with $h_i=0$ would lead to the requirement
     $s(k,u^*)=\infty$ which is impossible for finite $u^*$.

     So assume $u^*=\infty$. Then the finite $\beta_i$ condition implies that
     all $h_i$ are equal and at $u\to\infty$
\begin{equation}
     h_i u = \frac{2\lambda Cu +\omega +\psi}{A'(N+1)} +O(1). \label{Hiu}
\end{equation}
     On the other hand, the finite $\nu$ and $\beta$ conditions with
     (\ref{Hiu}) taken into account determine the asymptotics of $\omega$ and
     $\psi$ at $u\to\infty$:
\begin{eqnarray}
     \psi &=& -bu +O(1),\ \ \ b= \lambda C + ka^2/(N-1); \nn\\
     \omega &=& -(b+k)u +O(1)                               \label{AsOPsi}
\end{eqnarray}
     where we must put $k>0$ to obtain $\e^{\gamma} \to 0$.

     Now substitute the asymptotics (\ref{AsOPsi}) to (\ref{IntG}) with $h_i$
     and $C$ expressed in terms of $b$ and $k$ (provided $\lambda\ne 0$).
     After some algebra the left-hand side of (\ref{IntG}) takes the form
\begin{equation}
     b^2/\lambda^2 + o(1)                                   \label{BLam}
\end{equation}
     since by (\ref{AsOPsi}) $\psi'=-b + o(1),\ \omega' = -(b+k)+o(1)$.

     Eqs. (\ref{AsOPsi}) also imply $\psi''=o(1),\ \ \omega''=o(1)$.
     Hence it follows from (\ref{Eomega},\ref{Epsi}) that both $\e^{\psi}$
        and $\e^{\omega}$ tend to zero at $u\to\infty$, which is possible only
        if $b>0$. We arrive at a {\sl contradiction}: the left-hand side of
        (\ref{IntG}) tends to a positive value (by (\ref{BLam})) while its
     right-hand side tends to zero.

     The so far omitted case $\lambda=0$ is simple: from (\ref{AsOPsi}) we
     immediately obtain $b=0$ which is, as before, incompatible with
     $\e^{\psi}\to 0$ as required by the asymptotics of Eqs.
     (\ref{Epsi},\ref{Eomega}) with nonzero $q$ and $q'$.

     There is still one more possibility, namely, that
     the map with the coordinate $u$ is
     incomplete in the present static frame of reference. This may happen if
     $u=\infty$ is a regular surface and another coordinate must be used to
     penetrate beyond it, where anything may be found, including a horizon.
     However, this possibility is not realized in our system. Indeed, assume
     that $\beta,\ \beta_i,\ \nu$ and $\gamma$ are finite at $u=\infty$. Then
     by (\ref{BetaiG}) all $h_i=0$, hence $B'=0$ and from (\ref{BetaG}) it
     follows that $s(k,u)$ must have a finite limit at $u\to\infty$,
     contrary to the definition (\ref{Def-s}) of $s(k,u)$.

     Thus a BH-horizon is inconsistent with $q$ and $q'$ being
     simultaneously nonzero.  The same is true for a T-horizon as well since
     our equations are symmetric with respect to the functions $\gamma$ and
     $\nu$. (The only asymmetry, the possibility of $\eta_v=-1$, is
        insignificant for the above argument).  The theorem is proved.
$\bullet$

\section{String metric}                                       

     Let us briefly discuss the properties of Solutions A, B, C from Sect. 3.
     As pointed out in the Introduction, in the case $\lambda=\lst=
     pm \sqrt{N+1}$ it is more adequate to study the field behavior
     in terms of the so-called string metric
\begin{equation}
     \g_{MN}=\e^{-2\lambda\varphi}g_{MN}                         \label{MapS}
\end{equation}
     rather than the ``Einstein
     metric'' $g_{MN}$ which is most convenient for solving the equations.
     For other $\lambda$ the same can be done by analogy, leading to the
     action (\ref{Action}) in terms of $\g_{MN}$
\begin{equation}
     S= \int d^D x \sqrt{\g^D}\e^{(N+1)\lambda \varphi}\Big[\hat{R}^D
     +[1-\lambda^2(N+1)(N+2)]\g^{MN}\varphi_{,M}\varphi_{,N}
               -\hat{F}^2\Big]                              \label{ActS}
\end{equation}
     where symbols with hats denote quantities obtained with or corresponding
     to $\g_{MN}$.

     Nevertheless, we will discuss only the case $\lambda=\lst$ for which the
     metric $\g_{MN}$ is manifestly meaningful. Moreover, as we saw in Sect.
     3, $\lambda=\lst$ is the condition under which Solution C exists.

        Noteworthy the metric $\g_{MN}$ in all the solutions under study looks
        considerably simpler than $g_{MN}$.

\vskip1ex

     {\bf Solution A}

\nopagebreak
\vskip0.7ex

     With (\ref{PhiA}) the solution has the form
\begin{eqnarray}
     \ds_D^2 &=& \e^{2\omega}dt^2 -
 \e^{-2\C u}\left\{\frac{\e^{-2Bu}}{s^2 (k,u)}
     \left[\frac{du^2}{s^2 (k,u)}+d\Omega^2\right]
     -\sumi \e^{2h_i u}ds_i^2\right\},                   \label{DsAS}\\
 \lambda\varphi &=&
 \frac{N\C u - \omega}{N+1};\ \ \ F^1_{ut}= -q\e^{2\omega} \label{FAS}
\end{eqnarray}
     where $\omega$ is given by (\ref{OC}) and $\C = \lst C$.

     The metric behaves in a unique way if $h<0$ and/or $u_1<0$ when
     $0 < u < \umx < \infty$ ($u=0$ is the spatial infinity), $\umx$ being
     the smallest positive zero of the function $s(h, u+u_1)$. Namely, the
     only metric coefficient that becomes singular at $u\to\umx$ is
     $\g_{tt}\to\infty$, forming a repelling RN-type singularity (notably
     such a behavior is possible only provided $q\ne 0$). All other
     $\g_{MN}$ are finite and smooth everywhere, including the
     sphere $u=\umx$, whereas the fields $\varphi$ and $F$ are singular.
     Unlike the RN solution and its scalar-field counterpart in 4 dimensions,
     as well as the Einstein metric (\ref{DsA}), here the singularity occurs
     at a sphere of finite radius.

     If $h\geq 0,\ u_1>0$, then $k\geq 0$ as well and the solution is defined
     for all $u>0$. In the limit $u\to\infty$ the coefficients by $dt^2$,
     $d\Omega^2$ and $ds_i^2$ are, respectively,
\begin{equation}
     \g_{tt}=\e^{2\omega}\to 0;\ \ \g_{22}\approx \e^{-2(\C+B+k)u};\ \
          \g_{ii}\approx \e^{2(h_i-\C-B)u}.                  \label{LimAS}
\end{equation}
     Thus only the behavior of $\g_{tt}$ is unambiguous, that of the other
     $\g_{MN}$ strongly depends on the choice of the constants $\C$ and
     $h_i$, for instance, $\g_{22}$ and /or some $\g_{ii}$ may tend to
     infinity. However, if we require that both $\g_{22}$ and all $\g_{ii}$
     be finite at $u\to\infty$, we immediately arrive at the conditins
     (\ref{ConBH}) giving the BH solution.

     The string version of the metric (\ref{BH}) is
\begin{equation}
     \ds_D^2 = \frac{(1-2k/R)dt^2}{(1+p/R)^2}
          -\frac{dR^2}{1-2k/R} - R^2 d\Omega^2 + \sumi ds_i^2  \label{BHS}
\end{equation}
     while the conformal factor connecting $\g_{MN}$ and $g_{MN}$ is
\begin{equation}
     \e^{-2\lambda\varphi}= (1+p/R)^{-2/(N+1)}              \label{Confac}
\end{equation}
        Noteworthy, in this string BH
     metric all extra differents are ``frozen'' and exert no influence on the
     4-dimensional physics.

     The 4-dimensional version of (\ref{BH}) has been recently widely
     discussed (\cite{garf,gibmaeda,banks}, etc.). However, results of
        greater interest have been obtained with the magnetic-field version of
        (\ref{BH}) for which (in the case $D=4$) the metric is the same but the
        field $\varphi$ is replaced by $\overline{\varphi}=-\varphi$. So the
        conformal factor in (\ref{MapS}) is just $(1+p/R)$ and the resulting
        metric is in the notation (\ref{r+})
\begin{equation}
        \ds_4^2 = \frac{1-r/r_+}{1-r/r_-}dt^2 -
                        \frac{dr^2}{(1-r_+)(1-r_-)} -r^2 d\Omega^2.
\label{BHSM}
\end{equation}
        In the extreme case $k=0,\ r_+ = r_-$ the limit $r\to r_+$ corresponds
        to an infinitely long regular tube (``horn''). Such extreme
        dilatonic BH's (actually not BH's since there is no
        horizon in this case) are considered as the possible final state of
        BH evaporation \cite{banks}.

\vskip1ex

     {\bf Solution B}
\nopagebreak
\vskip0.7ex

     The conformal factor has the same form (\ref{Confac}) although
        $p$ can now have either sign.
     The general form of the solution for $\lambda=\lst$ is
\begin{eqnarray}
     \ds_d^2 &=& \eta_v\e^{2\psi}dv^2+ \e^{-2\C u}\left\{
 \e^{2h_0 u}dt^2 - \frac{e^{-2Bu}}{s^2(k,u)}\bigg[\frac{du^2}{s^2(k,u)}\bigg]
     + \summ \e^{2h_i u}ds_i^2 \right\}                     \label{DsBS}\\
        \lambda\varphi &=&
 \frac{N\C u - \psi}{N+1};\ \ \ F^1_{uv}= -q'\e^{2\psi}     \label{FBS}
\end{eqnarray}
     where $\psi$ is given in (\ref{PsiC}) and again $\C = \lst C$.

     The metric behavior is analysed similarly to Solution A. In particular,
     for $h'<0$ and/or $u_1<0$ the whole 4-dimensional metric remains regular
     at $u \geq \umx \infty$, while $\g_{vv}$ and the fields $\varphi, F$ are
     singular at the sphere of finite redius $u=\umx$.

     Although the 4-metric is regular at the surface $u=\umx$, one can hardly
     imagine that matter or field can safely cross it since a singularity
     must show itself in some way, such as runaway particle production or
     classical instability, as described in Sect. 8.

     In the T-hole case, realized under the conditions (\ref{ConBH}) ($h\to
     h'$) the string metric is
\begin{equation}
     \ds_D^2 =  \frac{(1-2k/R)\eta_v dv^2}{(1+p/R)^2} +
               dt^2- \frac{dR^2}{1-2k/R}
          - R^2 d\Omega^2 + \summ ds_i^2                    \label{THS}
\end{equation}
     The field $\varphi$
     is determined from (\ref{Confac}) and $F$ from (\ref{TH}).

     If $\eta_v=1$, the surface $R=2k$ is a Schwarzschild-like horizon in the
     $(R,v)$ subspace. However, if some points on the $v$ axis are
     identified, as should be done to compactify $V_1$ in the conventional
     way, then the corresponding sectors are cut out in the Kruskal picture,
        so that the $T$-domain and $R$-domain sectors join each other only in a
        single point, namely, the horizon intersection point.

     A similar thing happens if $\eta_v=-1$. Thus, in the domain $R>2k$ the
     compactified spatial coordinate $v$ is naturally described as an angular
     one ($0\leq v < 2\pi l$, where $v=0$ and $v=2\pi l$ are identified and
     $l$ is the compactification radius at the asymptotic $R\to\infty$). Then
     $R=2k$ is just the center of symmetry in the $(R,v)$ subspace; the
     latter has the shape of a tube having a constant thickness at
     $R\to\infty$, becoming narrower at smaller $R$ and and ending at $R=2k$
     either smoothly (if the regular center condition $l=2(2k+p)$ is
     satisfied), or with a top-of-a-cone singularity (otherwise). This
     suggests that there is no way to go beyond $r-2k$. On the other hand,
     it is unclear why an inhabitant of the physical space moving along the
     radius should turn back at the regular sphere $R=2k$ as if reflected
     from a wall.

     From the mathematical standpoint, this ambiguity may be
     resolved by just postulating that or other manifold structure underlying
     the metric. However, it would be better to physically justify such a
     postulate. The stability study of Sect. 8 may be treated as an attempt
     to find such a justification.

\vskip1ex
     {\bf Solution C}
\nopagebreak
\vskip0.7ex

     The above conformal mapping with $\varphi$ from (\ref{PhiC}), applied to
     the metric (\ref{DsC}) with $d=1$ under the condition $\lambda=\lst$,
     gives:
\begin{equation}
  \ds_D^2 = \eta_v\e^{2\psi}dv^2 + \e^{2\omega}dt^2
     +\e^{-2C'u/(N-1)}
     \bigg\{\frac{\e^{-B'u}}{s^2 (k,u)} \bigg[\frac{du^2}{s^2 (k,u)}
          +d\Omega_{d+1}^2\bigg] +\summ \e^{2h_i u} ds_i^2 \bigg\}
                                                                 \label{DsCS}
\end{equation}
     where $\omega$ and $\psi$ are described by (\ref{OC}) and (\ref{PsiC}),
     while $F^1_{ut}$ and $F^2_{uv}$ are determined by the corresponding
     formulas in (\ref{FAS}) and (\ref{FBS}).

        From (\ref{IntC}) (with $d=1$), it is not hard to conclude that if
     $k<0$, then at least one of the constants $h$ or $h'$ is ``more
     negative'' than $k$: $|h|\ ({\rm or}\ |h'|)\ \geq |k|$. Therefore, in all
     cases when $0<u<\umx <\infty$ the quantity $\umx$ coincides with the
     smallest positive zero of one of the functions $\e^{-\omega}$ or
     $\e^{-\psi}$. Thus at $u=\umx$ either $\g_{vv}$, or $\g_{tt}$, or both
     tend to infinity, while other $\g_{MN}$ remain regular (a modified RN
     singularity).

     If $\umx = \infty$, the metric behavior is diverse, depending on
     the interplay of the integration constants. It seems unreasonable to
     enumerate all the possibilities.
     The most significant conclusion is that there is a naked singularity in
     all cases; horizons are absent according to Theorem 1.

     For the case of arbitrary $\lambda$, when exact solutions are hard to
     obtain, another general statement can be proved in addition to Theorem
     1:

\medskip
{\bf Theorem 2.} {\sl The field system  (\ref{Action},
        \ref{Stru},\ref{DsD}) cannot form a
     static, spherically symmetric configuration with a regular center.}
\medskip

     A {\sl proof} makes use of the same set of equations as that of Theorem
        1 and a similar type of argument. $\bullet$
\medskip

{\sl Remark.} The regular center conditions include the regularity
     requirements for $g_{MN}$ and $\varphi$. Consequently, as is easily
     shown, the statement of Theorem 2 is valid as well for the string metric
     $\g_{MN}$ and all other conformal gauges connected with $g_{MN}$ by
     factors of the form $\exp(\const\cdot\varphi)$. The same is true for
     Theorem 1 if the (natural) additional requirement is adopted that
     $\varphi$ should be finite at a horizon.

     Regular spherical configuration with no center at all,
     like wormholes or ``cornucopions'' \cite{banks}
     are certainly not excluded, as seen from the above consideration.

\section{$D$-dimensional solutions from the 4-dimensional viewpoint}  

     A 4-dimensional version of (\ref{Action}) is obtained by integrating out
     the extra-dimension coordinates, so that (up to a constant factor and
     a divergence)
\begin{equation}
     S=\int d^4 x \sqrt{^4 g}
     \e^{\sigma}\left[^4 R-\sigma^{,\mu}\sigma_{,\mu}
     +\sumi N_i\beta_{i,\mu}\beta_i{}^{,\mu} + \varphi^{,\mu}\varphi_{,\mu}
          -\e^{2\lambda\varphi} F^{1\mu\nu}F^1_{\mu\nu}
     -2\eta\e^{-2\nu+2\lambda\varphi}W^{,\mu}W_{,\mu} \right]  \label{Act4}
\end{equation}
     where Greek indices run over the values 0, 1, 2, 3, $^4 R$ is the
     curvature derived from the 4-dimensional part $g_{\mu\nu}$ of $g_{MN}$
     and $\sigma= \sumi N_i\beta_i$. It is assumed that
     $F^1= dW^1,\ W^1 = U_{\mu}(x^{\alpha})dx^{\mu}$ and
     $F^2= dW^2,\ W^2 = W(x^{\alpha})dv$ where, as before, $v$ parametrizes
     the 1-dimensional subspace $V_1$ and $\nu=\beta_1$. Thus $F^1$ is the
     electromagnetic field and $F^2$ forms a new scalar field $W(x^\alpha)$,
     both coupled to $\varphi$ and all $\beta_i$. The field $W$ is minimally
     coupled to $g_{\mu\nu}$ and the sign of its kinetic term depends on
     $\eta_v$: it is normal if $v$ is spacelike and anomalous if $v$ is
     timelike.

     The action (\ref{Act4}) corresponds to the $D$-Einstein conformal gauge.
     The 4-dimensional Einstein gauge with the metric $\gg_{\mu\nu}$
     is obtained by the conformal mapping similar to that used by Dicke
        \cite{dicke} in his scalar-tensor theory and by Wagoner \cite{wag}
        for more general scalar-tensor theories
\begin{equation}
     \gg_{\mu\nu}= \e^{\sigma}g_{\mu\nu}             \label{Map4E}
\end{equation}
     after which the action takes the form
\begin{equation}
     S=\int d^4 x \sqrt{^4 \gg}
     \left[^4 \overline{R}+ \half\sigma^{,\mu}\sigma_{,\mu}
     +\sumi N_i\beta_{i,\mu}\beta_i{}^{,\mu} + \varphi^{,\mu}\varphi_{,\mu}
          -\e^{\sigma+2\lambda\varphi} F^{1\mu\nu}F^1_{\mu\nu}
     -2\eta\e^{-2\nu+2\lambda\varphi}W^{,\mu}W_{,\mu} \right]  \label{Act4E}
\end{equation}
     where indices are raised and lowered using $\gg_{\mu\nu}$.

     The actions (\ref{Action}) and (\ref{Act4E}) are equally convenient for
     solving the field equations due to the constant effective gravitational
     coupling. Noteworthy that the coordinate $u$ as introduced in
        (\ref{Harm}) is harmonic with respect to both the $D$-dimensional
metric
        $g_{MN}$ and the 4-dimensional metric $\gg_{\mu\nu}$, i.e., $\nabla^M
        \nabla_M u = \overline{\nabla}^{\mu}\overline{\nabla}_{\mu} u =0$, but
     not with respect to the 4-dimensional part $g_{\mu\nu}$ of the
     $D$-metric:  $\nabla^{\mu}\nabla_{\mu} u \ne 0$.

     The metric $\gg_{\mu\nu}$ thus corresponds to the so-called
     {\sl gravitational system of measurement} \cite{dicke,stanmel}. However,
        real space-time measurements, such as solar-system experiments,
        rest on the constancy of atomic quantities (the {\sl atomic system of
        measurements}). Thus, the modern definition of reference length is
        connected with a certain spectral line, determined essentially by the
        Rydberg constant and ultimately by the electron and nucleon masses.
        Therefore observational properties of various theoretical models, like
        multidimensional configurations, are most reasonably described in a
        conformal gauge where masses of bodies of nongravitational matter, such
        as atomic particles, do not change from point to point.

        In other words, in the gauge to be selected (to be denoted
        $g^*_{\mu\nu}$) the nongravitational matter Lagrangian $L_m$ should
        enter into the action with no $\sigma$- or $\varphi$-dependent factor
        \cite{rad}. However, the choice of $g^*_{\mu\nu}$ depends on how $L_m$
        appears in the original action, that is, how matter is coupled to the
        metric and dilaton fields in the underlying fundamental theory.

        In \cite{green}, where the effective field-theoretic limit of string
        theory in 10 dimensions is given in a form similar to (\ref{Action})
        (Eq. (13.1.49)) some quadratic fermion terms do not contain the dilaton
        field. If those terms are associated with matter, then in our
simplified
        model it is reasonable to write $L_m$ just as an additional term in the
        brackets of Eq. (\ref{Action}).Then, passing over to the 4-dimensional
        formulation, it is easy to check that the metric in the ``atomic
gauge''
        should have the form
\begin{equation}
      g^*_{\mu\nu}  = \e^{\sigma/2}g_{\mu\nu}              \label{MapA}
\end{equation}

        The term $L_m$ would enter (\ref{Act4}) and (\ref{Act4E}) with the
        factors $\e^{\sigma}$ and $\e^{-\sigma}$, respectively, whereas in
        terms of $g^*_{\mu\nu}$ the corresponding part of the action is just
        $\int d^4 x\sqrt{g^*} L_m$. The same metric $g^*_{\mu\nu}$ would be
        obtained if we wrote the action for  a point particle moving in $D$
        dimensions in the conventional form $-\int m\, ds $ and required that
in
        the 4-dimensional metric $g^*_{\mu\nu}$ it move along geodesics
        \cite{rad}.

        The notion of active gravitating mass of an isolated object in a space
        with the structure (\ref{Stru}) is to be also introduced with the aid
of
     $g^*_{\mu\nu}$, by comparing $g^*_{tt}$ far from the source with the
        expression $(1-2GM/r)$ of the Schwarzschild metric, so that, with an
        arbitrary radial coordinate $u$
\begin{equation}
     GM= -|g^*_{\theta\theta}|^{3/2}\partial_u{g^*_{tt}}/
               \partial_u g^*_{\theta\theta}\Big\mid_{u\to u_{\infty}}
       = r^2{\gamma^*}'/r'\Big\mid_{u\to u_{\infty}}             \label{Mass}
\end{equation}
     where $\e^{2\gamma^*}=g^*_{tt},\
        r^2 = -g^*_{\theta\theta}$ ($\theta$ being the polar angle in the
        spherical coordinate system) and $u_{\infty}$ is the value of $u$ where
        $r\to\infty$ and $\gamma^*\to 0$.

        Specifically, for dilatonic BH's (\ref{BH})
\begin{equation}
     GM = k+ p\frac{3N+1}{4(N+a^2)},                         \label{BHM}
\end{equation}
        while for T-holes (\ref{TH})
\begin{equation}
     GM = \frac{k}{4} - \frac{p}{2(a^2+N)}                 \label{THM}
\end{equation}
        For other solutions $M$ can be easily found as well using (\ref{MapA})
        and (\ref{Mass}).

        Eq. (\ref{BHM}) generalizes the corresponding relation for dilatonic
        BH's in 4 dimensions ($D=4,\ N=1$). For instance (see also
        (\ref{r+})), extreme BH's, those with the greatest charge for given
mass
        or the smallest mass for given charge correspond to $k=0$; at $N=1$
this
        happens for $GM=|Q|/(1+a^2)$. At $k\to 0$ the horizon is squeezed to a
     point (the center) and becomes a singularity.

     Similarly to the RN case,
     at greater charges for given mass naked singularities exist.

        As for T-holes, Eq. (\ref{THM}) shows that these objects in our model
     may have negative gravitational masses, i.e., be felt by test particles
     as repellers. That happens if $\eta_v = +1$ for large $Q'$ (see
     (\ref{TH}).

     In conformal gauges other than (\ref{MapA}) test particle trajectories
        $x^j (t)\ (j=1,2,3)$ are certianly the same but they are no longer
        geodesics: particles move under joint action of the metric and scalar
     fields, just like the situation discussed by Dicke \cite{dicke}.

\section{Stability analysis}

        The stability of Solution A under small perturbations preserving
        spherical symmetry was studied in Refs.
        \cite{potsdam,br-vuz92,bm-itogi}. The system with two dynamic degrees
of
        freedom (the dilaton field and a single extra-dimension scale factor)
        was considered and three cases when the perturbation equations decouple
        were studied in detail. It was concluded that solutions with naked
        singularities are catastrophically unstable, while the BH ones are
     stable. These results generalized the earlier ones from \cite{brohod}
     and \cite{brokir} where the instability of static
     scalar-(electro)vacuum configurations in conventional general
     relativity was established, in particular, of BH's with scalar charge.

     Our main interest here is to determine the properties of
        T-holes, therefore let us study a special case of Solution B with $D=5\
        (N=2)$, so that there are 3 dynamic degrees of freedom, connected with
        $\varphi,\ \nu$ and the gauge-field component $W$.

        Thus we are considering a 5-dimensional system with the metric
\begin{equation}
        ds_5^2 = \e^{2\gamma}dt^2 - \e^{2\alpha}du^2
                - \e^{2\beta}d\Omega^2 + \eta_v \e^{2\nu}dv^2   \label{Ds5P}
\end{equation}
     where $\alpha,\beta,\gamma,\nu$, as well as  $\varphi$ and the gauge
     field potential $W$ differ from those in the static Solution B of Sect.
     3 only by small (linear) deflections depending on $u$ and $t$
\begin{equation}
        \dg,\ \da,\ \delta\beta,\ \dn,\ \df,\ \dW               \label{Perturb}
\end{equation}

     The perturbation problem is easier to consider in the 4-dimensional
     Einstein conformal gauge (\ref{Map4E},\ref{Act4E}) where in our case
     $\sigma= \beta_1 =\nu$ and $F^1=0$.

     The next step is to choose the frame of reference and the coordinates
     in the perturbed space-time. This choice may be carried out by
     prescribing certain relations among the perturbations. Following
     \cite{brohod}, we would like to choose the so-called central frame of
     reference, that where coordinate spheres of fixed radii are at rest, and
     the radial coordinate such that the numerical values of these radii are
     the same as those in the static background configuration with the metric
     $\gg_{\mu\nu}$. Thus we postulate $\delta\rho\equiv 0$ where
     $\rho\equiv\ln |\gg_{\theta\theta}| = 2\beta+\nu$.

\def\Ro{\overline{R}}
\def\alphao{\overline{\alpha}}
\def\gammao{\overline{\gamma}}

     The nonzero Ricci tensor components corresponding to
     $\gg_{\mu\nu}$ are
\begin{eqnarray}
     \Ro^t_t &=& \e^{-2\gammao}\ddot{\alphao}-\e^{-2\alphao}\big[\gammao''
          + \gammao'(\gammao' + \rho' - \alphao')\big];   \nn\\
     \Ro^u_u &=& \e^{-2\gammao}\ddot{\alphao}-\e^{-2\alphao}\big[\gammao''
          + \rho'' -\alphao'(\gammao'+\rho')-{\gammao'}^2-{\rho'}^2/2\big];
                                                          \nn\\
     \Ro^{\theta}_{\theta} = \Ro^{\phi}_{\phi} &=&
          \e^{-\rho} - \half\e^{-2\alphao}
               \big[\rho''+ \rho'(\gammao'+\rho'-\alphao'\big]; \nn\\
     \Ro_{ut} &=& -\rho'\dot{\alphao}.                      \label{RicciP}
\end{eqnarray}
     where $\e^{2\gammao}=\gg_{tt},\ \e^{2\alphao}=-\gg_{uu}$, dots and
     primes denote $\partial_t$ and $\partial_u$, respectively, where the
     coordinate $u$ is harmonic in the static background solution in both
     metrics $g_{\mu\nu}$ ($\alpha= 2\beta+\gamma+\nu$) and $\gg_{\mu\nu}$
     ($\alphao = \rho+ \gammao$); in the perturbed solution $u$ is no longer
     harmonic.

     The field equations and the energy-momentum tensor components
     corresponding to $\nu,\ \varphi$ and $W$ are found directly from
     (\ref{Act4E}). Singling out the perturbations and taking into account
     the equations for the background static solution, one obtains the
     following coupled linear wave equations for $\df,\ \dn$ and $\dW$:
\begin{eqnarray}
     \L\df + \varphi'(\da'-\dg') &=& 4\lambda\eta_v q'(\dW'-W' \dps);
                                                            \label{Edphi}\\
     \L\dn + \nu'(\da'-\dg') &=& -(8/3)q'(\dW'-W' \dps);  \label{Ednu}\\
     \L\dW + W'(\da'-\dg') &=& -2W' \dps                  \label{EdW}
\end{eqnarray}
     where the operator $\L = \e^{2\rho}\partial_t^2- \partial_u^2,\
     \dps=\dn-\lambda\dg$ and all coefficients by the perturbations are taken
     from the background solution. The combination
     $\da' - \dg'\equiv \delta\alphao' - \delta\gammao'$ is determined from
     the  $(^\theta_\theta)$ component of the Einstein equations giving
\begin{equation}
     \da'-\dg' = 4\delta\alphao\e^{2\alphao-\rho}\big/ \rho'.  \label{Diff}
\end{equation}
     Here $\delta\alphao$ is in turn expressed in terms of the dynamic
     variables by integrating the mixed $(ut)$ component of the Einstein
     equations in $t$:
\begin{equation}
     2\rho'\delta\alphao = 3\nu'\dn + 2\varphi'\df -4\eta_v q'\dW +F(u)
                                                           \label{Dalpha}
\end{equation}
     where $F$ is an arbitrary function. Rejecting the static perturbations,
     i.e., putting $F\equiv 0$, and making use of the relation
     $\e^{2\alphao-\rho}=\rho''$ valid for the static solution, we obtain the
     expression to be substituted to the wave equations in the form
\begin{equation}
     \da'-\dg' = (\rho''/{\rho'}^2)(3\nu'\dn+2\varphi'\df-4\eta_v q'\dW).
                                                                \label{Diff'}
\end{equation}

     Our static system is unstable if there exist
     physically allowed solutions to the wave equations
     growing at $t\to \infty$. A separate problem is to define which
     solutions should be accepted as physically allowed ones. Let us join the
     approach of \cite{brohod} and require
\begin{equation}
     \dn \to 0, \ \
     \df \to 0, \ \ \dW \to 0 \ \ {\rm for} \ \ \  u\to 0   \label{Infty}
\end{equation}
     at spatial infinity $r\to \infty$ and
\begin{equation}
     \mid \dn/\nu \mid < \infty,
     \ \ \mid \df/\varphi \mid <\infty,
     \ \ \mid \dW/W\mid <\infty                             \label{Horsing}
\end{equation}
     at singularities and horizons. These requirements may be called
     minimal since they just provide the validity of linear perturbation
     theory over the whole space-time, including the neiborhood of the
     singularities. (If some more restrictive boundary conditions can be
     physically grounded, our instability conclusions must be revised.) In
     addition, we ought to forbid energy fluxes to our system from
     outside; however, similarly to \cite{brohod},  that would lead to just a
     constraint on the signs of the integration constants without
     affecting any of the further conclusions.

     The wave equations are rather complicated:
     it seems hard not only to solve but even to decouple them in the general
     case.  Therefore let us consider here only the simplest case $q'=0$, the
     purely scalar-vacuum 5-dimensional background solution described by a
     special case of (\ref{GTan}):
\begin{eqnarray}
     ds_5^2 &=& f^{a_0}dt^2 -f^{-a_0-a_1}dR^2
          - f^{1-a_0-a_1}R^2d\Omega^2 + \eta_v f^{a_1}dv^2;    \nn \\
     \e^{2k\varphi}&=& f^{\CC};\ \ \
               a_0^2 + a_0a_1 + a_1^2 + \CC^2 =1   \label{SVac5}
\end{eqnarray}
     with $f(R) = 1-2k/R$. The BH and T-hole cases are, respectively,
     $a_0=1,\ a_1=\CC=0$ and $a_0=\CC=0,\ a_1=1$.

     If $q'=0$, then both $\nu'$ and $\varphi'$
     are constants and Eqs. (\ref{Edphi}-\ref{EdW}) decouple:
\begin{eqnarray}
     \L\df_1 = (3{\nu'}^2 + 2{\varphi'}^2)\df_1;\ \ \L\df_2=\L\df_3=0;
                                                            \nn \\
     \df_1 \equiv 3\nu'\dn + 2\varphi'\df;\ \ \df_2\equiv\varphi'\dn-\nu'\df;
      \ \ \ \df_3\equiv \dW                                   \label{Ewave}
\end{eqnarray}
     Separating the variables in (\ref{Ewave}) and transforming $\mu$ and $u$
     to obtain the normal Liouville form of the mode equations,
\begin{equation}
     \df_j = \e^{i\Omega_j t -\rho/2}y_j(x), \ \ x=-\int \e^{\rho(u)}du
\end{equation}
     $(j=1,2,3)$, we obtain the Schr\"{o}dinger-like equations
\begin{equation}
     d^2 y_j/dy^2 +[\Omega_j^2-V_j(x)]y_j=0                   \label{Schrod}
\end{equation}
     with the effective potentials
\begin{equation}
     V_1=V_3=-\e^{-3\rho/2}(\e^{\rho/2})'';\ \
          V_2= V_1+4k^2(1-c^2)\e^{-2\rho}(1/\rho')'             \label{Vu}
\end{equation}
     where $c= a_0+a_1/2$. These are the expressions for $V_j$ in terms of
     the harmonic coordinate $u$; however, they can be presented explicitly
     in terms of the $R$ coordinate of (\ref{SVac5}):
\begin{equation}
     V_1=V_3= \frac{k[2cR-k(1+c)^2]}{R^{2+2c}(R-2k)^{2-2c}};
     \ \ \ V_2=V_1 - \frac{2(1-c)^2}{R^{1+2c}(R-2k)^{1-2c}}.    \label{VR}
\end{equation}
     The variable $x=\int (1-2k/R)^{-c}dR$ behaves like $R$ at large $R$ and
     tends to either $-\infty$ (for $c=1$, the BH case), or to a
     finite value, say, 0 (otherwise).

     In the BH case $V_1=V_2=V_3 > 0$ at $R>2k$.
     The boundary conditions (\ref{Infty},\ref{Horsing}) correspond to those
     conventional in quantum mechanics and thus the positiveness of $V_j$
     means that solutions to (\ref{Schrod}) with $\Omega_j^2<0$ are absent.
     Consequently, multidimensional BHs are stable under linear monopole
     perturbations.

     If $c<1$, then at $R\to 2k$
\begin{equation}
     x\to 0, \ \ V_1=V_3\approx V_2= -\frac{1}{4x^2}(1+o(1)). \label{Xto0}
\end{equation}
     since the difference $V_2 - V_1$ is negligible as compared with $V_1$.
     Thus $V_j$ have a negative pole at $x=0$ and tend to zero at
     $x\to\infty$.

     The variable $W$ is absent in the background configuration, as is
     $\varphi$ in special cases of it. Therefore one should require
     $|\dW| < \infty$ and $|\df| < \infty$ at $x\to 0$. However, there is no
     reason to require $|\dn| < \infty$ since $\nu \to -\infty$ at $R\to 2k$.
     Therefore the boundary conditions for $\df_1$ and $\df_2$ take the form
\begin{eqnarray}
     y_j\to 0 \ \ \ && {\rm at}\ x \to \infty,               \label{YInfty}\\
     |y_j|/(\sqrt{x}\ln x)<\infty && {\rm at}\ x \to 0.      \label{Yzero}
\end{eqnarray}
     The asymptotic forms of the solutions to Eqs. (\ref{Schrod}) for
     $\Omega_j^2<0$ at $x\to 0$ and $x\to \infty$ are
\begin{eqnarray}
     x\to\infty:\ y_j&=& C_{1j} \e^{-|\Omega_j|x} +C_{2j} \e^{|\Omega_j|x},
                                                            \label{Yas}\\
     x\to 0:\ \ \ y_j&=& \sqrt{x}(C_{3j} + C_{4j}\ln x).        \label{Ysing}
\end{eqnarray}
     Evidently the condition (\ref{Horsing}) is satisfied for any $C_{3j}$ and
     $C_{4j}$.  Therefore the solutions $y_j(x)$ chosen on the basis of
     (\ref{Infty}), i.e., the ones with $C_{2j}=0$, are admissible and
        realize the instability of our static system.

     In other words, in the potential well corresponding to the naked
     singularity there exist arbitrarily low ``energy levels'' for the
     perturbations.  Thus the system is unstable and the instability is of a
     catastrophic nature since the increment $|\Omega|$ has no upper bound.

     The T-hole case ($c=1/2$) is here not distinguished among the singular
     configurations corresponding to other values of $c$. So the instability
     conclusion is valid for T-holes on equal grounds with them.

     The general conclusion is that all the above configurations, except
     BH's, are unstable.

     The results of this section confirm those of Refs.
     \cite{potsdam,br-vuz92,bm-itogi} on the stability of BH's and the
     instability of other (singular) configurations. In some respects the
     cited results are extended here, since previously there have been no
     stability study of (\ref{SVac5}) with $\varphi\ne 0$ and no study of
     perturbations in the form of the gauge field $\dW$.

\section{Concluding remarks}                                     

\noindent {\bf 9.1.}
     Having considered a large variety of spherically symmetric solutions for
     our field system, forming $(n+3)$- and even $(n+4)$-parameter families,
     we nevertheless have to conclude, by the stability criterion, that
     just the BH family with only two free parameters, to be identified with
     mass and charge, and two input constants, the space-time dimension and
     the dilatonic coupling constant, can be called realistic. This explains
     why, for instance, the possibility of a nonzero gauge field component in
     an extra dimension can be ignored, etc., and extends our previous
     results \cite{potsdam,br-vuz92,bm-itogi}.
 \vskip 1ex

\noindent {\bf 9.2.}
     We have seen that T-holes, the possible windows to
     space-time domains with unusual physics, appear as solutions to
     multidimensional field equations as frequently as do BH's.

     As in the present state of the Universe extra dimensions are generally
     beleived to be compactified to a very small size, it is hard to imagine
     how T-holes might now form from ordinary matter. However, in the early
     Universe where all dimensions could be equally relevant, T-holes could
     form on equal grounds with primordial BH's and consequently their
     relics might play a certain role at the present stage, e.g., be one
     of the forms of dark matter. However, such a possibility is questionable
     due to the instability of these objects.

     Multidimensional objects as possibly forming dark matter were
     recently discussed by Wesson \cite{wesson} who suggested such a
     possibility for purely vacuum 5-dimensional ``solitons'' \cite{davow}
     coinciding with the special case $\CC=0$ of (\ref{SVac5}).
     Notably it is the simplest T-hole (\ref{TSchw}) if we put, in
     addition, $a_0=0, a_1=1$. We have shown that the solutions considered by
     Wesson are strongly unstable under monopole
     perturbations, therefore his particular example of multidimensional dark
     matter candidate looks unlucky.

\vskip 1ex
\noindent {\bf 9.3.}
     Only uncharged ($q'=0$) T-holes have been explicitly shown to be
     unstable. However, very probably it is the case for charged ones as
     well, since perturbations near a T-horizon behave just as they do near a
     singularity, while all singular solutions studied so far
     \cite{potsdam,br-vuz92,bm-itogi} turned out to be unstable.
     Nevertheless, T-hole solutions in other field models, which
     exist for sure, may in principle turn out to be stable, although, on the
     other hand, it may happen that there is a kind of ``censorship'' like
     Hawking's chronology protection conjecture \cite{hawk}, confirmed, in
        particular, by the instability of Cauchy horizons.

\vskip 1ex
\noindent {\bf 9.4.}
     Certain difficulties in the T-hole description arise due to the
     compactification of extra dimensions. However, the latter may be
     invisible to 3-dimensional observers for a reason other than their small
     size, such as, e.g., the behavior of field potentials, as discussed in
     Ref. \cite{rubshap} (we may live within a 3-dimensional ``membrane'' at
     the bottom of a potential ``trench'' in a multidimensional world).
     A possible T-hole existence in such models may be a subject of interest
     for further study.

\vskip 1ex
\noindent {\bf 9.5.}
     As seen from Sect.6, there is a significant difference between
     electrically and magnetically charged configurations in the presence of
     a dilaton field and/or extra dimensions, unlike conventional general
     relativity. The RN metric describes on equal grounds
     electrically, magnetically and dually charged BH's (dyons). In
     our case, with $\varphi\ne 0$, the 4-dimensional metrics for electric
     and magnetic charges coincide in the Einstein conformal gauge but are
     different in the string gauge. An electric charge is compatible
     with any hyperspherical ($O(d+2)$) symmetry of the space-time, whereas a
     magnetic charge (at least, defined in the conventional way) can
     exist only within the ordinary spherical ($O(3)$) symmetry. In the
     general multidimensional case the difference is still more significant.

     In a subsequent paper (in preparation) it will be shown that
     static solutions of magnetic monopole and dyon types (the latter if
     $\lambda= \lst$) can be obtained for our system (\ref{Action}) in the
     general case (\ref{Stru}); although the solution procedure is
     essentially the same as here, the results are different.

\vskip4ex

\centerline{\bf Acknowledgement}
\vskip2ex   \nopagebreak
     This work was in part supported by the Russian Ministry of Science.

        It is a pleasure to thank my colleagues V.Iva\-shchuk,
        M.Kon\-stan\-ti\-nov, V.Mel\-ni\-kov, A.Ra\-dy\-nov and P.R.L.V.Moniz
        for useful discussions.

\end{document}